\documentclass[twocolumn,floatfix,groupedaddress,superscriptaddress,aps,pra,showpacs]{revtex4-1}

\usepackage{graphicx}
\usepackage{amsmath}
\usepackage{amssymb}
\usepackage{amsfonts}
\usepackage{dcolumn}
\usepackage{bbm}
\usepackage{float}

\usepackage{color}
\newcommand{\operator}[1]{\hat{#1}}
\newcommand{\vektor}[1]{\textbf{#1}}

\newcommand{\iu}{\mathrm{i}}

\begin{document}

\title{Topologically nontrivial Hofstadter bands on the kagome lattice}

\author{Christoph H.\ Redder}
%\email{xx.yy@tu-dortmund.de}
\affiliation{Lehrstuhl f\"{u}r Theoretische Physik I, 
Technische Universit\"{a}t Dortmund,
 Otto-Hahn Stra\ss{}e 4, 44221 Dortmund, Germany}

\author{G\"otz S.\ Uhrig}
\email{goetz.uhrig@tu-dortmund.de}
\affiliation{Lehrstuhl f\"{u}r Theoretische Physik I, 
Technische Universit\"{a}t Dortmund,
 Otto-Hahn Stra\ss{}e 4, 44221 Dortmund, Germany}

\date{\textrm{\today}}

\begin{abstract} 
We investigate how the multiple bands of fermions on a crystal lattice
evolve if a magnetic field is added which does not increase the number
of bands. The kagome lattice is studied as generic example for
a lattice with loops of three bonds. Finite Chern numbers occur as a nontrivial topological property in the presence of the magnetic field. 
The symmetries and periodicities as a function of the applied field
are discussed. Strikingly, the dispersions of the edge states
depend crucially on the precise shape of the boundary.
This suggests that suitable design of the boundaries helps
to tune physical properties which may even differ between 
upper and lower edges. Moreover, we suggest a promising gauge to
realize this model in optical lattices.
\end{abstract}


% 03.65.Vf, Phases: geometric; dynamic or topological
% 03.75.Lm, Tunneling, Josephson effect, Bose-Einstein condensates in periodic potentials, solitons, vortices, and topological excitations (see also 74.50.+r Tunneling phenomena; Josephson effects in superconductivity)
% 02.40.Pc 	General topology
% Lattice fermion models, 71.10.Fd

\maketitle

\section{Introduction}

Since the discovery of the quantum Hall effects \cite{klitz80,tsui82},
interest in topological aspects of condensed-matter systems
has risen and has stayed very high ever since. The link to
topological invariants is given by the Berry phase \cite{berry84} of the
ground-state wave function in its dependence of magnetic fluxes
on a torus \cite{niu85,wegne88}. This line of argument 
even shows that there are no corrections to Ohm's law on the linear
relationship between voltage and current \cite{klein90,uhrig91}.

Only a few years ago, the
discovery of topological insulators \cite{ando13} gave another impetus 
to the field of topology in condensed-matter physics.
Even in classical mechanical systems topological edges states
can be excited \cite{susst15}.
Very recently, topologically nontrivial bands were realized
in lattices of ultracold atoms \cite{jimen12,aidel15} and advocated in
strongly frustrated spin systems with Dzyaloshinskii-Moriya interactions
\cite{romha15}.

For the present article, we are inspired by the realization of
nontrivial gauge fields in optical lattices filled by ultracold atoms
\cite{jimen12,aidel15} and by the interest in nontrivial lattices.
Whereas Aidelsburger \textit{et al.}\ realized bands with Chern numbers different
from zero in Bravais lattices, we show that similar physics
also occurs in crystal lattices, i.e., lattices with a basis.
Moreover, our focus is on lattices with loops of odd numbers of bonds.
The generic loop is a triangle which has the smallest odd number of bonds.
In antiferromagnetic spin systems, it is the prime source of frustration.

In our theoretical study, we aim at a proof-of-principle result. Though inspired
by the impressive recent advances in systems of ultracold atoms in optical lattices 
\cite{jimen12,aidel15,jotzu14},
we do not claim that this class of systems provides the most promising
candidate for experimental realizations. The obstacle may be that the traps
constructed so far do not provide well-defined edges, which we examine below.
But solid-state systems may fill the gap. The recent observation of the
quantum anomalous Hall effect in thin layers for ferromagnetic Chern insulators
provides seminal progress \cite{chang13,kou14}. Presently, higher temperatures
are reached at which the effect occurs \cite{chang15} and theoretical calculations even suggest
that in tailored systems experiments at room temperature are possible \cite{wu14,han15}.
There are concrete suggestions for tailored superlattices which should also
make the design of particular edges possible \cite{krash11,han15}. An alternative 
to solid-state lattices may be the artificial lattices built from dots or antidots
in tailored semiconductor structures \cite{lan12}.

\begin{figure}[htb]
\begin{center}
	\includegraphics[width=0.9\columnwidth,clip]{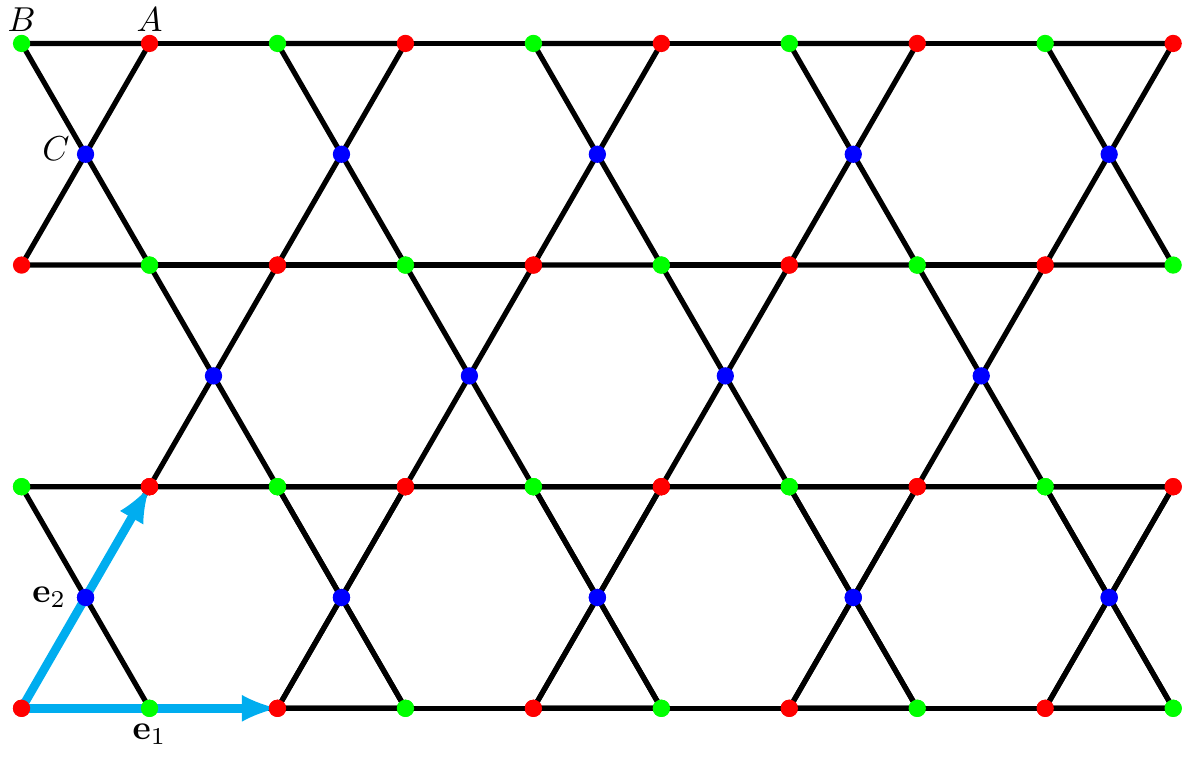}
\end{center}
\caption{(Color online) 
Sketch of the kagome lattice, which is a crystal lattice
with a basis of three sites (A, B, and C). Also shown is one
choice of primitive vectors $\vektor{e}_i$ (in light blue) 
which span the unit cell.
\label{fig:kagome}
}
\end{figure}

In the present fundamental investigation, 
we study the kagome lattice shown in Fig.\ \ref{fig:kagome}.
This crystal lattice has a basis of three atoms, labeled A, B, and C
in the sketch. One choice of primitive vectors is shown; they span
a parallelogram which constitutes a unit cell. Note that the coordination 
number is only $z=4$ which is rather low in comparison to $z=6$
for the triangular lattice, the generic Bravais lattice with triangular loops. This lattice has appeared before 
in several studies.
In particular, we show that the kagome lattice in
certain magnetic fields corresponds to special cases of Haldane
models \cite{halda88b}, i.e., models without
uniform magnetic field but complex hoppings, on the kagome lattice 
\cite{ohgus00,katsu10a,tang11}. Moreover, in the research field 
investigating flat band models and interaction effects on them
the kagome lattice is also studied intensively; for a review see
Ref.\ \cite{bergh13}.

%%%%%%%%%%%%%%%%%%%%%%%%%%%%%%%%%%%%%
%%%%%%%%%%%%%%%%%%%%%%%%%%%%%%%%%%%%%%%%%%%%%%%%%%%%%%%%%%%%%%%%%%%%%%%%%%%%%%%

We study this model for the special magnetic fields where 
the number of bands still equals the number of bands without
magnetic field. Note that this is an important difference
from the investigation of the square lattice by Aidelsburger \textit{et al.}\
where topological effects could only arise once the number of
bands was increased by certain values of the magnetic field \cite{aidel15}.
The bands and their dispersion are computed as well as the first Chern 
number $C_n$. This Chern number
 equals the Berry phase \cite{berry84} occurring for a Bloch
state which surrounds the Brillouin zone. The Brillouin zone is
the relevant two-dimensional manifold, namely, a simple torus $T^2$.
This Chern number is defined by
\begin{subequations}
\label{eq:chernzahl}
\begin{eqnarray}
&& C_n  := \frac{1}{2\pi\iu} \int_{T^2} F^{(n)} df \quad \in \mathbb{Z}
\label{eq:curvature1}
\\
 && = \frac{1}{2\pi\iu} \int_{T^2} 
(\langle \partial_1 u_{n\vektor{k}}|\partial_2 u_{n\vektor{k}}\rangle
- \langle \partial_2 u_{n\vektor{k}}|\partial_1 u_{n\vektor{k}}\rangle)
dk_1dk_2 . \qquad
\label{eq:curvature2}
\end{eqnarray}
\end{subequations}
In Eq. \eqref{eq:curvature1} $F^{(n)}$ is the Berry curvature \cite{berry84}  
of the principal $U(1)$ fiber bundle defined by the $n$th Bloch eigenstate over the Brillouin zone. 
Equation\ \eqref{eq:curvature2} is the explicit formula in terms of
the Bloch states $|u_{n\vektor{k}}\rangle$. The partial derivatives $\partial_i$
refer to the derivation with respect to the momenta $k_i$, i.e.,
$\partial_i:=\partial/\partial k_i$. 
The above Chern number must be an integer since 
it can be converted by the Stokes theorem to the integrated phase along
a closed path divided by $2\pi$.

We compute the above Chern number for the three kagome bands,
thereby identifying the topologically nontrivial bands.

\section{Model and magnetic fields}

For simplicity, we study a
nearest-neighbor tight-binding model on the kagome lattice.
Its Hamiltonian reads
\begin{equation}
\operator{H} = \sum_{\langle ij\rangle} t_{ij}c^\dag_{i}c_{j} + \sum_i V_i c^\dag_i c_i 
\label{eq:tightbinding}
\end{equation}
where the indices run over all sites and $\langle ij\rangle$ stands for
nearest neighbors. The creation operator $c^\dag_{i}$ creates a fermion
at site $i$ and $c_{i}$ annihilates it.
In general, we consider complex hopping elements $t_{ij}=t\exp(i\vartheta_{ij})$
with some directed phases $\vartheta_{ij}$ resulting from the Peierls substitution,
i.e.,
\begin{equation}
\vartheta_{ij} = \frac{q}{\hbar}\int_i^j\vektor{A}\,d\vektor{r},
\label{eq:peierls}
\end{equation}
where $q$ is the charge of the hopping particle.
The local potentials are given by $V_i$. They depend on the sublattice
to which the site $i$ belongs.

%%%%%%%%%%%%%%%%%%%%%%%%%%%%%%%%%%%%%
\begin{figure}[htb]
\begin{center}
	\includegraphics[width=1.0\columnwidth,clip]{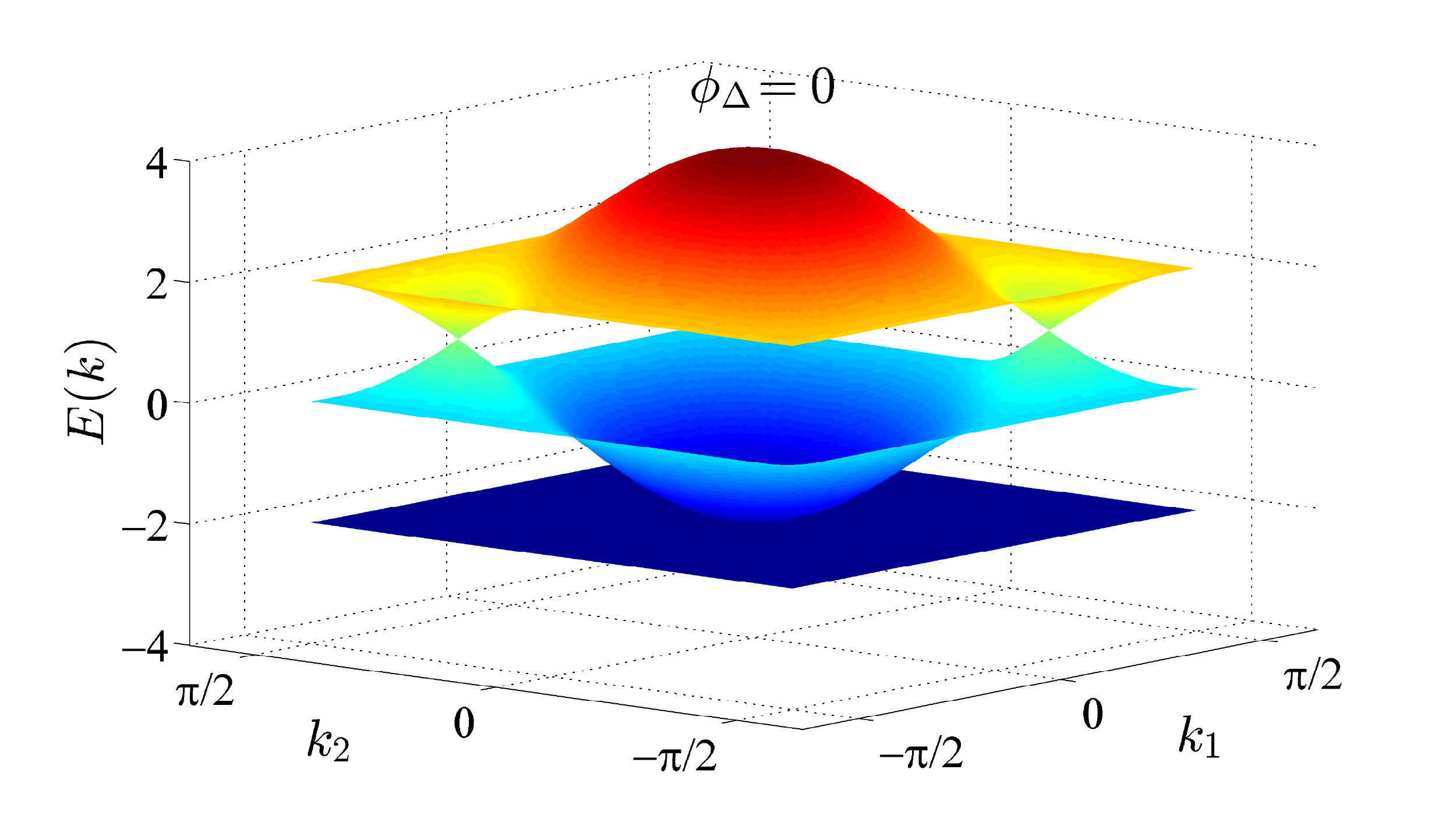}
\end{center}
\caption{(Color online) 
Dispersion  of the three bands as functions of $k_1$ and $k_2$
as defined in Eq.\ \eqref{eq:some_defs} in the kagome lattice without 
any magnetic field and local potentials. The hopping is set to $t=1$.
Note the flat lowest band
and the two Dirac cones where the two other bands touch with
linear dispersion.
\label{fig:flux_n0}}
\end{figure}
%%%%%%%%%%%%%%%%%%%%%%%%%%%%%%%%%%%%%%%%%%%%%%%%%%%%%%%%%%%%%%%%%%%%%%%%%%%%%%%

As a reference, we first look at the case without any magnetic fields
and without local potentials such that $\vartheta_{ij}=0$ and $V_i=0$.
It is known that in this case
 a flat band occurs because the kagome lattice is a line graph, namely, the
line graph of the honeycomb lattice. Line graphs are generally 
known to have flat bands  \cite{mielk91a,mielk91b}. The explicit calculation illustrates the flat band
nicely in Fig.\ \ref{fig:flux_n0}. It is based on the $3\times 3$ matrix in momentum
space 
\begin{equation}
\mathcal{H}_\vektor{k}=2t
\begin{pmatrix}
	0    & \cos k_1  &    \cos k_2\\
	\cos k_1 & 0 & \cos k_3\\
	\cos k_2 & \cos k_3 & 0 \\
\end{pmatrix},
\label{eq:blochmatrix1}
\end{equation}
where we use the wave-vector components
\begin{subequations}
\label{eq:some_defs}
\begin{eqnarray}
k_i &:=& \vektor{k} \cdot \mathbf{\Delta}_i,\\
\mathbf{\Delta}_1 &:=& \vektor{e}_1/2,\\
\mathbf{\Delta}_2 &:=& \vektor{e}_2/2,\\
\mathbf{\Delta}_3 &:=& \mathbf{\Delta}_2 - \mathbf{\Delta}_1
\end{eqnarray}
\end{subequations}
with $\vektor{e}_1 = (1,0)^\dag$ and $\vektor{e}_2 = (1,\sqrt{3})^\dag/2$;
i.e., the lattice constant from one A site to the nearest A site is set to 
unity. We stress that the full Brillouin zone is given
by $k_i\in[-\pi/2,\pi/2]$ because the $\mathbf{\Delta}_i$ have
only half the length of the primitive vectors.
Note that $k_3=k_2-k_1$ holds according to the above definitions;
only two of the $k_i$ are independent.

The energy eigenvalues of the matrix \eqref{eq:blochmatrix1}
are
\begin{subequations}
\begin{eqnarray}
E_{1,2\vektor{k}} &=& t(1\pm \sqrt{1+8\cos k_1 \cos k_2 \cos k_3}\,),
\\
 E_{3\vektor{k}} &=& -2t.
\label{eq:triv_res}
\end{eqnarray}
\end{subequations}
These bands are degenerate at $\vektor{k} = (0,0)^\dag$ where the
flat band touches the lower dispersive band and at 
 $\vektor{k}_{\pm} = \pm (2\pi/3)(1,-\sqrt{3})^\dag$ 
where the two dispersive bands display Dirac cones.
Due to the degeneracy, the Chern numbers  $C_n$ 
of the bands are not properly defined. But even if one breaks
the degeneracy, for instance by infinitesimal local potentials,
the $C_n$ turn out to be trivial, i.e., zero, because
the model without finite phases is time-reversal invariant
which implies $C_n=0$.

\begin{figure}[htb]
\begin{center}
	\includegraphics[width=1.0\columnwidth,clip]{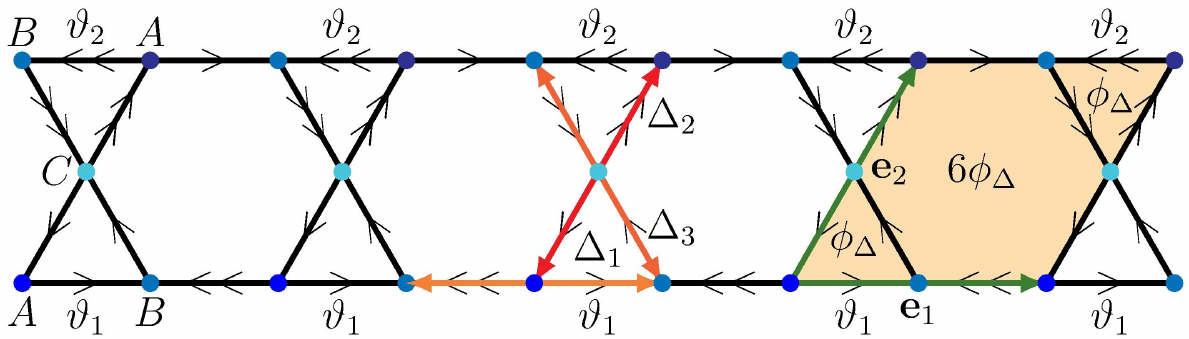}
\end{center}
\caption{(Color online) 
Gauge chosen for the specific calculations. Two  phases, 
$\vartheta_1$ and  $\vartheta_2$, are introduced which are directed
in the sense of the arrows.
\label{fig:phases}}
\end{figure}

Thus we turn to the model in a uniform magnetic field
perpendicular to the kagome plane. This induces finite phases
according to Eq.\ \eqref{eq:peierls}. We do not aim at discussing
the complete intricate interplay of discrete lattice symmetry and noncommuting
magnetic translations \cite{zak64a,zak64b,kohmo85} which give rise to 
multiple Hofstadter bands \cite{hofst76}. Instead, we aim at 
the situation where the magnetic translations $T_i$ along the
primitive vectors $\vektor{e}_i$ commute. 
They obey the relation $T_1 T_2=\exp(i\phi_\text{UC})T_2 T_1$
where $\phi_\text{UC}$ is the magnetic flux through the unit
cell measured in units of $\hbar/e$ \cite{zak64a,zak64b,kohmo85}.
Thus, in order not to increase the number of bands, we avoid
decreasing the translational symmetry. Thus we focus on
the commuting case $T_1 T_2=T_2 T_1$ implying 
\begin{equation}
\label{eq:quant1}
\phi_\text{UC}=2\pi n,
\end{equation}
where $n\in\mathbb{Z}$ is an arbitrary integer. Inspecting the unit cell shown
by the lightly shaded area in Fig.\ \ref{fig:phases} elementary
geometry shows that $\phi_\text{UC}=8\phi_\Delta$ if
$\phi_\Delta$ is the flux through a triangle. Thus Eq. \eqref{eq:quant1}
translates to 
\begin{equation}
\label{eq:quant2}
\phi_\Delta=\frac{\pi}{4} n.
\end{equation}
The physics modulo gauge transformations of the phases
is only influenced by the fluxes through the loops modulo
flux quanta, not by the individual phases. Thus we conclude that the physics, such as Chern numbers
of the bands, depends on $\phi_\Delta$ with periodicity of $2\pi$ so that
we will have to study only the values of $n$ modulo 8.

%%%%%%%%%%%%%%%%%%%%%%%%%%%%%%%%%%%%%
%%%%%%%%%%%%%%%%%%%%%%%%%%%%%%%%%%%%%%%%%%%%%%%%%%%%%%%%%%%%%%%%%%%%%%%%%%%%%%%

We choose a specific gauge where
the phases $\vartheta_{ij}$ take values as
depicted in Fig.\ \ref{fig:phases}. The arrows on the bonds
stand for the direction of the phases $\vartheta_1$ and $\vartheta_2$.
In this gauge, the total phase around the unit cell 
takes the value zero, which may be a surprise because it seems
to reflect zero flux and thus zero magnetic field. But this is not the case
because the sum of all phases along closed loops is meaningful
only modulo $2\pi$. Hence zero total phase complies with a finite
flux fulfilling Eq.\ \eqref{eq:quant1}. To make this point explicit we
consider a gauge for arbitrary flux in Appendix \ref{app:arbit} and show
that it can be regauged to the one in Fig.\ \ref{fig:phases}
in Appendix \ref{app:specific}. In this particular gauge,
the relation between these phases and the flux $\phi_\Delta$
is the following 
\begin{subequations}
\label{eq:micros_relation}
\begin{eqnarray}
3\vartheta_1 &=& \phi_\Delta - 2\pi\,m_1 , 
\\
3\vartheta_2 &=& \phi_\Delta - 2\pi\,m_2 ,
\\
-3(\vartheta_1 + \vartheta_2) &=& 6\phi_\Delta - 2\pi\,m_3 ,\\
0 & =&  8 \phi_\Delta - 2\pi\, n,
\label{eq:sum}
\end{eqnarray}
\end{subequations}
where  $m_i$ and $n\in \mathbb{Z}$ may occur since the fluxes 
through the loops are only fixed up to multiples of $2\pi$.
The last equation [Eq.\eqref{eq:sum}] is the sum of the three equations
before with $n=m_1+m_2+m_3$ which confirms Eq. \eqref{eq:quant1}.
The left-hand side of Eq. \eqref{eq:sum} stands for the 
vanishing sum of all phases around a unit cell; see Fig.\ 
\ref{fig:phases}. This does not mean that there is no
net uniform magnetic field through the lattice, but it reflects
the fact that the magnetic translations commute for
the special magnetic fields we are considering because 
the flux through the unit cell is a multiple of the flux
quantum. Thus, there are gauges without 
net total phase around the unit cells and we are
employing such a gauge here for simplicity.
But in the smaller loops, i.e.,
the triangles, there is a net flux given by $\phi_\Delta$.

For finite phases and local potentials we obtain the following
matrix problem:
\begin{equation}
\mathcal{H}_\vektor{k}=
\begin{pmatrix}
	V_A                                                  
	& 2te^{-\iu \bar{\vartheta}}c_1 & 2te^{\iu \bar{\vartheta}}c_2
	\\
	2te^{\iu \bar{\vartheta}}c_1  
	& V_B																							
	& 2te^{-\iu \bar{\vartheta}}c_3
	\\
	2te^{-\iu \bar{\vartheta}}c_2
	& 2te^{\iu \bar{\vartheta}}c_3
	&  V_C                                              
	\\
\end{pmatrix},
\label{eq:matrix2}
\end{equation}
where we use the shorthand
\begin{equation}
c_i:=\cos(k_i-(-1)^i \Delta\vartheta)
\end{equation}
for brevity. We have introduced the average and the difference
of the phases $\bar\vartheta:=(\vartheta_1+\vartheta_2)/2$
and $\Delta\vartheta:=(\vartheta_1-\vartheta_2)/2$. Obviously,
the matrix $\mathcal{H}_\vektor{k}$ is $2\pi$ periodic in 
$\bar\vartheta$ and $\Delta\vartheta$. Interestingly,
the band Hamiltonian \eqref{eq:matrix2} is identical to 
the ones considered in Haldane models before 
\cite{ohgus00,katsu10a,tang11} if $\Delta\vartheta=0$ and
the local potentials are switched off.
But we recall that Eq. \eqref{eq:matrix2} only holds
for uniform magnetic fields which meet the condition \eqref{eq:quant2}. This observation leads us to the experimentally useful result that certain Haldane models can be realized for specific
fluxes complying with Eq. \eqref{eq:quant2} by uniform magnetic fields.

But the eigenvalues
are even periodic in $\bar\vartheta$ with period $2\pi/3$. This
can be seen by the gauge transformation
\begin{equation}
\mathcal{U}=
\begin{pmatrix}
	1                                                  
	& 0 & 0
	\\
	0 	& 	e^{\iu 2\pi/3} & 0
	\\
	0	& 0	&   e^{\iu 4\pi/3}		                                           
	\\
\end{pmatrix},
\end{equation}
which transforms the Hamiltonian matrix according to
\begin{equation}
\mathcal{H}_\vektor{k}(\bar\vartheta + 2\pi/3)
= \mathcal{U} \mathcal{H}_\vektor{k}(\bar\vartheta) \mathcal{U}^\dag.
\end{equation}
The value of $\Delta\vartheta$ remains unchanged.
Clearly, the eigenvalues at each momentum do not change under
the transformation and hence they are $2\pi/3$-periodic 
in $\bar\vartheta$. Since the transformation $\mathcal{U}$
is independent of momentum it does not change the Chern numbers
either so that the Chern numbers are also $2\pi/3$-periodic.
Note that a change of $\bar\vartheta$ by $2\pi/3$ at constant 
$\Delta\vartheta$ corresponds to the simultaneous change 
\begin{equation}
\label{eq:gauge1}
m_1\to m_1-1, m_2\to m_2-1, \quad \text{and} \quad m_3\to m_3+2.
\end{equation}
Alternatively, the increase of $\bar\vartheta$
by $2\pi/3$ can be interpreted as an increase of $\phi_\Delta$
by $2\pi$ and the increment $m_3\to m_2+8$. Thus, we see
that the physics is indeed $2\pi$-periodic in $\phi_\Delta$.

In addition, one easily sees that $\bar\vartheta\to \bar\vartheta+\pi$
simply leads to a global minus sign in the Hamiltonian matrix
\eqref{eq:matrix2}, thus inverting the sequence of bands and 
thereby swapping the Chern numbers of the first and the third bands.
The change $\bar\vartheta\to \bar\vartheta+\pi$ directly corresponds to
the change $\phi_\Delta \to \phi_\Delta + 3\pi$; see Eq.\ 
\eqref{eq:micros_relation}. But in view of the $2\pi$-periodicity
in $\phi_\Delta$
we find the corresponding bands up to a gauge transformation 
already for $\phi_\Delta \to \phi_\Delta + \pi$.
For examples, we refer the reader to the Sec.\ 
\ref{sec:results} below.

Moreover, reversing time amounts up to changing
the sign of the phases $\vartheta_i \to -\vartheta_i$ and
of the flux $\phi_\Delta\to-\phi_\Delta$; i.e., the Hamiltonian
matrix is complex conjugated. This transformation
of $\mathcal{H}_\vektor{k}$ leaves the eigenvalues and,  
thus the dispersion unchanged because the energy eigenvalues 
are real numbers. But it changes the sign of the Chern numbers. 

Combined with the shift of the flux by $\pi$
we obtain the result that the change
$\phi_\Delta \to \pi-\phi_\Delta$ inverts the energy bands, 
i.e., takes the dispersion energy to their negative values,
but leaves the Chern numbers unchanged.
We will illustrate these properties in Sec.\ 
\ref{sec:results} below.

The Chern numbers of the bands are also $2\pi/3$-periodic
in $\Delta\vartheta$ because the shift of the momenta
\begin{subequations}
\label{eq:shift}
\begin{eqnarray}
k_1\to k_1 + 2\pi/3,\\
k_2\to k_2 - 2\pi/3,
\end{eqnarray}
\end{subequations}
implies $k_3\to k_3 - 4\pi/3 = k_3 + 2\pi/3$,
where the last identity holds modulo $2\pi$. Thus,
this shift of the momenta corresponds to the change of
gauge $\Delta\vartheta\to \Delta\vartheta + 2\pi/3$.
So the eigenvalues at given momenta change, but they
are only shifted in reciprocal space. Since the Chern
number is a property of the band in the Brillouin zone
it does not change under Eqs. \eqref{eq:shift}.
This is obvious if one uses the shifted Brillouin zone,
shifted by the same amount as in Eqs. \eqref{eq:shift}.
Note that a change of $\Delta\vartheta$ by $2\pi/3$ at constant 
$\bar\vartheta$ corresponds to the simultaneous change 
\begin{equation}
\label{eq:gauge2}
m_1\to m_1+1, m_2\to m_2-1, \quad \text{and} \quad m_3\to m_3.
\end{equation}

We stress that the combination of the two gauge transformations,
Eqs. \eqref{eq:gauge1} and
\eqref{eq:gauge2}, illustrates that the essential physics is determined
by $n=m_1+m_2+m_3$ while the individual values of the $m_i$ 
do no matter much.
However, $m_3$ cannot be regauged to change by unity only.

\section{Results}

\label{sec:results}

In this section, we show explicit results for the dispersion
and the Chern numbers of the three bands of the kagome lattice
in a magnetic field obeying relation \eqref{eq:quant2}.

\subsection{Results without local potentials}

For vanishing local potential the eigenvalues of the matrix \eqref{eq:matrix2}
can be determined analytically. Solving the characteristic polynomial
of order 3 we obtain
\begin{subequations}
\begin{eqnarray}
E_{b\vektor{k}} &=& 4t \sqrt{{Q}} 
\cos\left([\theta_\vektor{k}+2\pi b]/3\right), \quad b \in \{0,1,2\},
\\
\theta_\vektor{k} &: = &\arg\left(P\cos(3\bar{\vartheta})+
\iu\sqrt{Q^3-[P\cos(3\bar{\vartheta})]^2}\right),	\qquad
\\
P&:=& c_1 c_2 c_3,\\
Q&:=& (c_1^2+c_2^2+c_3^2)/3.
\end{eqnarray}
\end{subequations}

%%%%%%%%%%%%%%%%%%%%%%%%%%%%%%%%%%%%%
\begin{figure}[htb]
\begin{center}
	\includegraphics[width=1.0\columnwidth,clip]{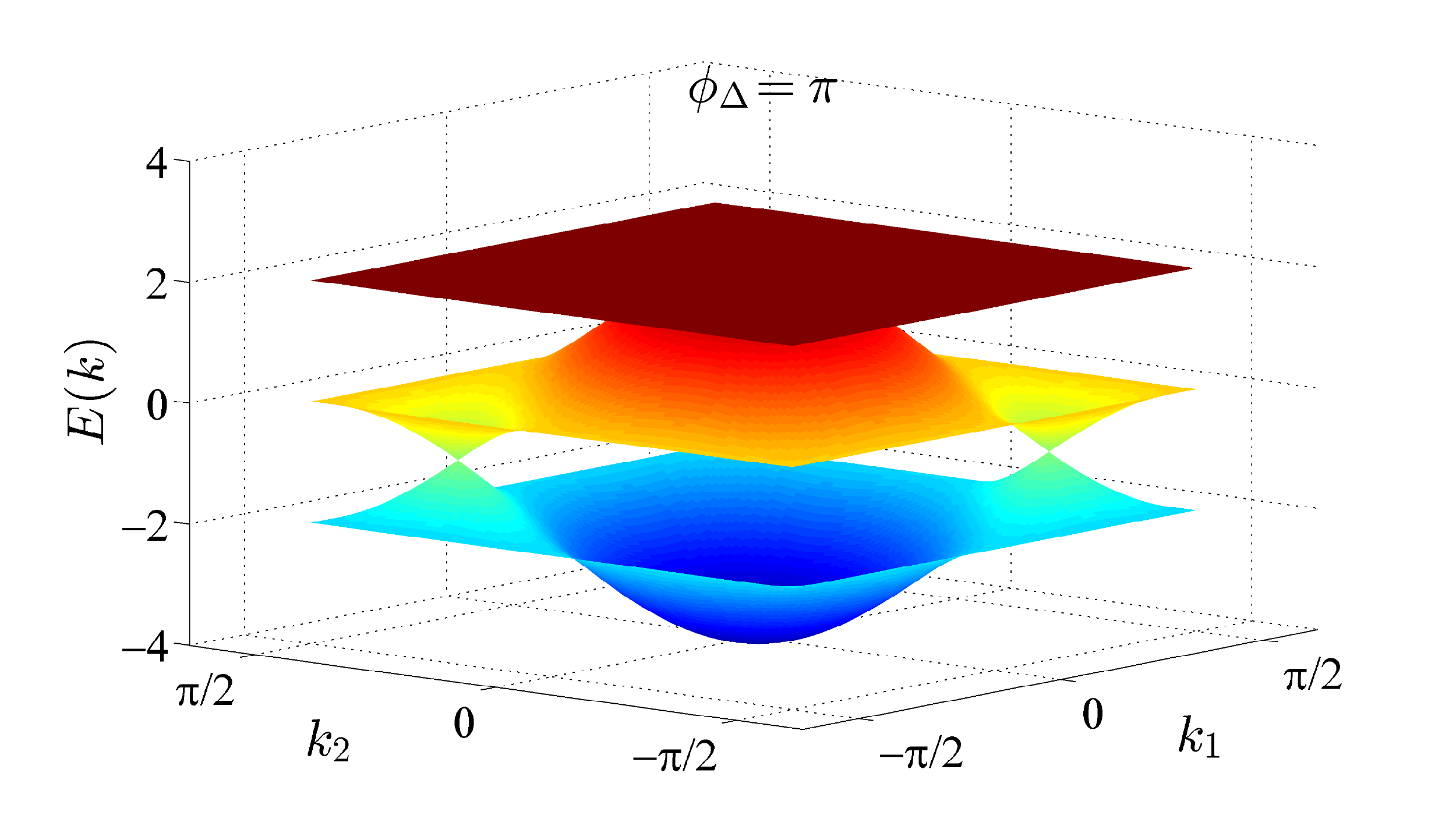}
\end{center}
\caption{(Color online) 
Dispersion  of the three bands as functions of $k_1$ and $k_2$
as defined in Eq.\ \eqref{eq:some_defs} in the kagome lattice 
at $\phi_\Delta=\pi$, i.e., $n=4$ in Eq. \eqref{eq:quant2}, 
and $\vartheta_i=\phi_\Delta/3$ and
no local potentials. The hopping is set to $t=1$.
\label{fig:flux_n4}}
\end{figure}
%%%%%%%%%%%%%%%%%%%%%%%%%%%%%%%%%%%%%%%%%%%%%%%%%%%%%%%%%%%%%%%%%%%%%%%%%%%%%%%

For simplicity, we choose $m_1=m_2=0$, thus $\Delta\vartheta=0$, and 
vary $m_3=n$ to achieve various fluxes $\phi_\Delta$. As pointed out
above, one only has to consider eight values of $\phi_\Delta$ because
of the $2\pi$-periodicity and the quantization condition
\eqref{eq:quant2}. Additionally,
$\phi_\Delta\to \phi_\Delta+\pi$ only inverts the energies and
swaps Chern numbers so that it is sufficient to consider $n=0,1,2,3$.
For instance, $n=4$ corresponds to adding the phase $\pi$ and, indeed,
Fig.\ \ref{fig:flux_n4} displays the negative bands of
Fig.\ \ref{fig:flux_n0}. The degenerate, touching
bands do not allow for an unambiguous definition of the Chern number.
But the fact that the Hamiltonian matrix is equivalent to 
a real matrix up to a gauge transformation \eqref{eq:gauge1}
teaches us that the bands are topologically trivial $C_n=0$.

%%%%%%%%%%%%%%%%%%%%%%%%%%%%%%%%%%%%%
\begin{figure}[htb]
\begin{center}
	\includegraphics[width=1.0\columnwidth,clip]{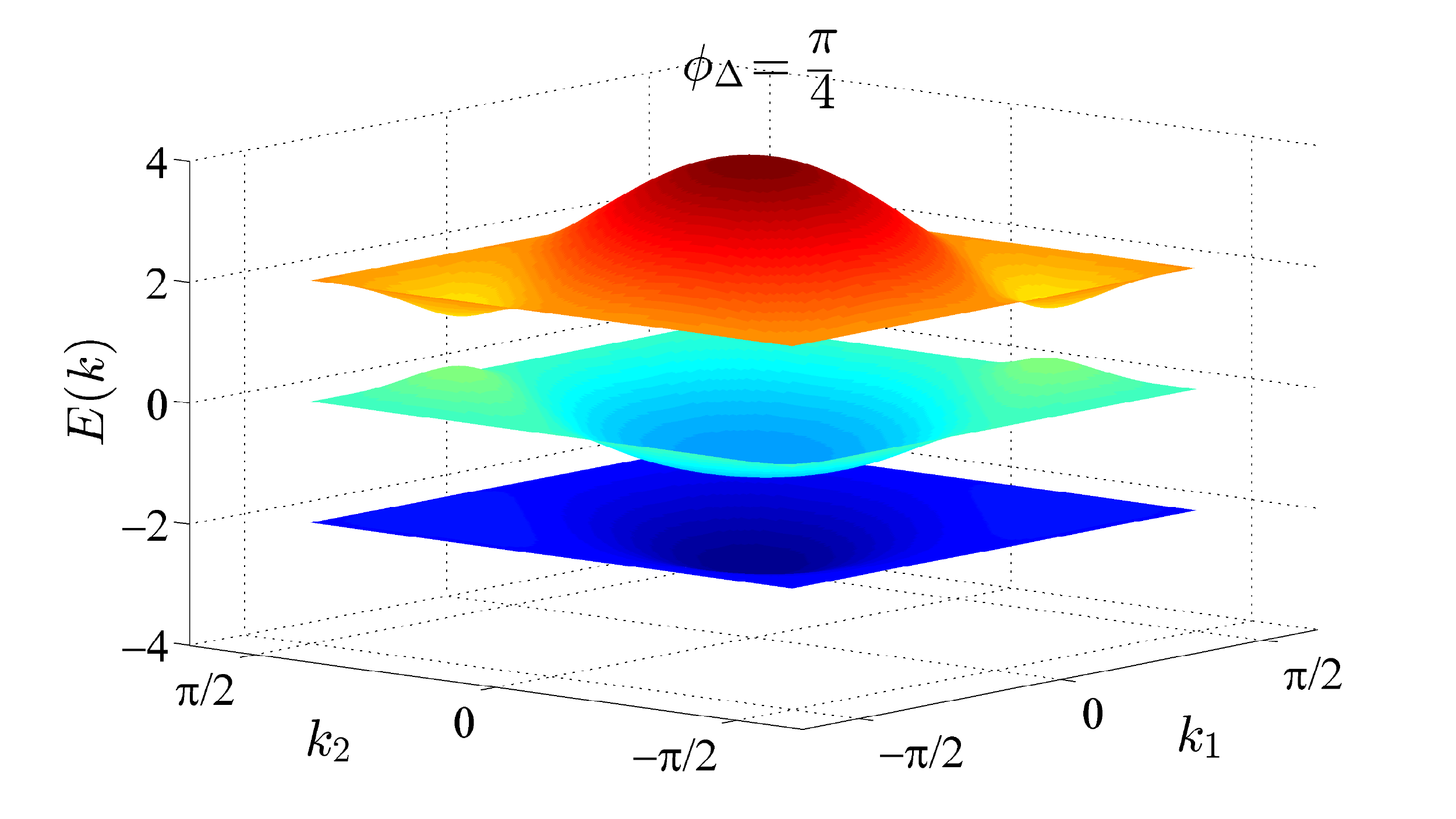}
	\includegraphics[width=1.0\columnwidth,clip]{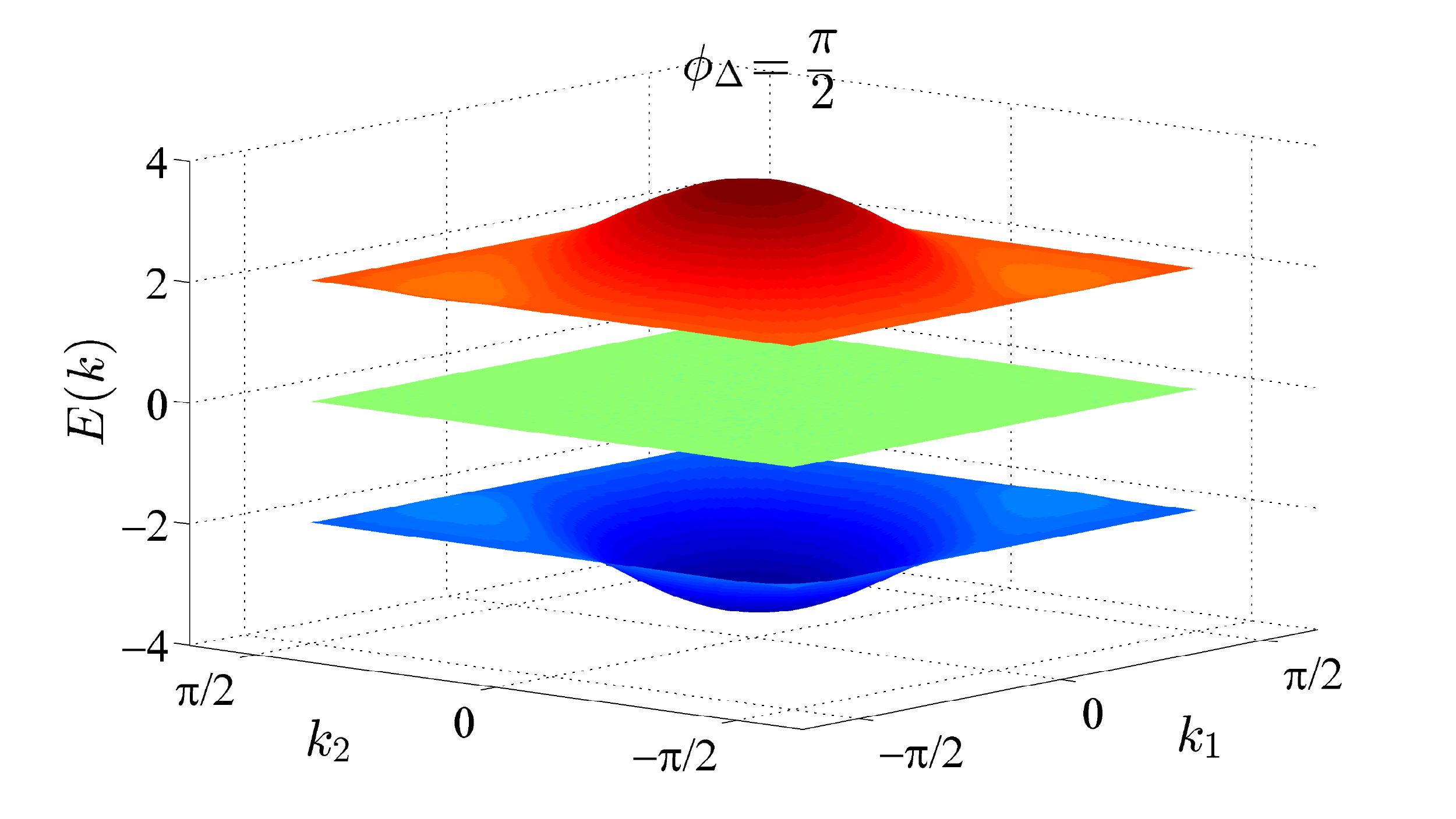}
	\includegraphics[width=1.0\columnwidth,clip]{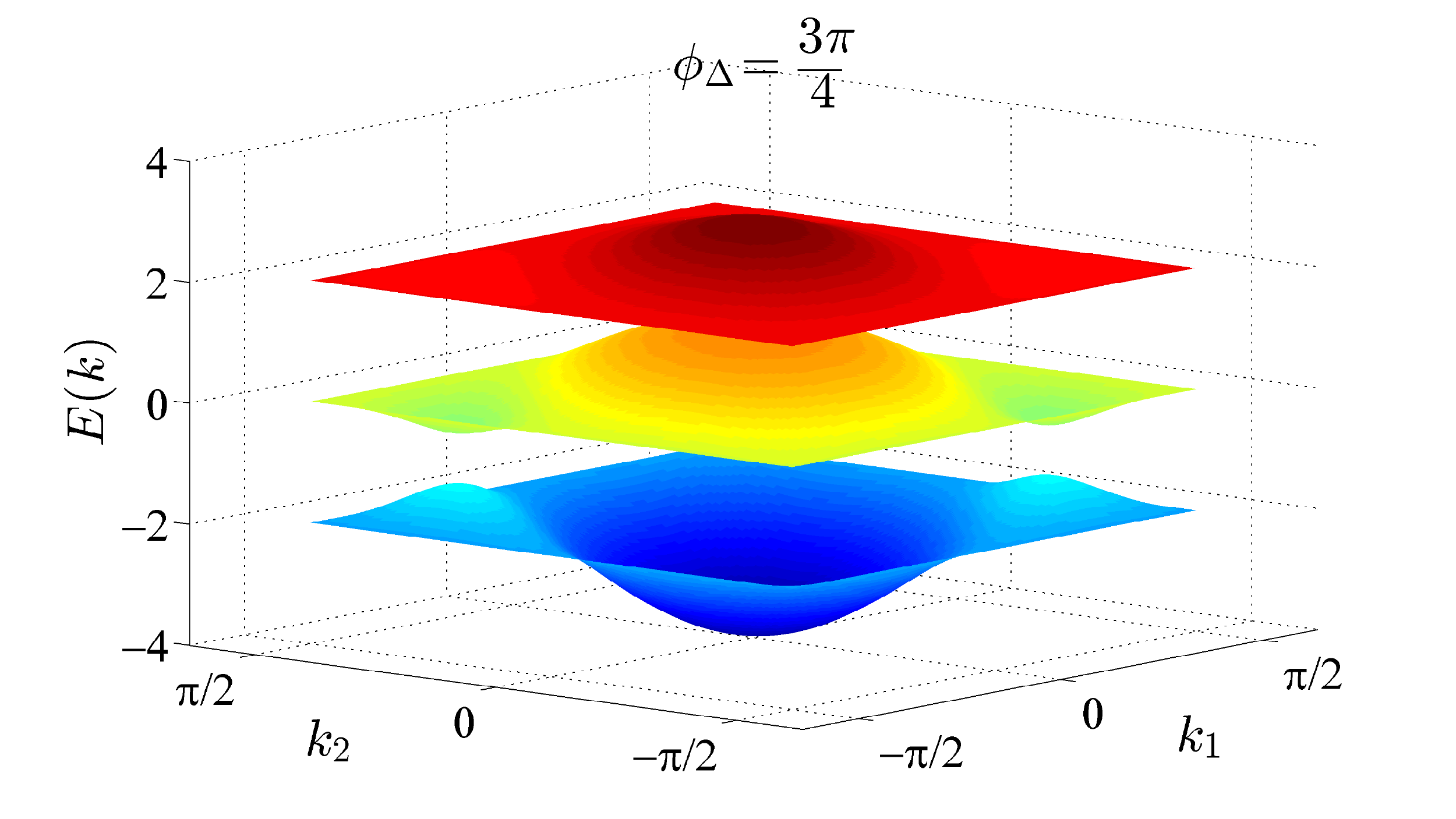}
\end{center}
\caption{(Color online) 
Dispersion  of the three bands as functions of $k_1$ and $k_2$
as defined in Eq.\ \eqref{eq:some_defs} in the kagome lattice 
at $\phi_\Delta=(\pi/4)n$ for $n=1$ (top), 
$n=2$ (middle), and $n=3$ (bottom)
and $\vartheta_i=\phi_\Delta/3$ and
no local potentials. The hopping is set to $t=1$.
\label{fig:flux_n123}}
\end{figure}
%%%%%%%%%%%%%%%%%%%%%%%%%%%%%%%%%%%%%%%%%%%%%%%%%%%%%%%%%%%%%%%%%%%%%%%%%%%%%%%

The three cases $n=1,2,3$ are much more interesting; 
the corresponding bands are depicted in
Fig.\ \ref{fig:flux_n123} in ascending order.
All three bands are separated from one another
so that the Chern numbers are well defined.
Note the gradual evolution upon increasing 
flux, illustrated nicely from Fig.\ \ref{fig:flux_n0}
over the panels of Fig.\ \ref{fig:flux_n123}
to Fig.\ \ref{fig:flux_n4}.

The fact that $\phi_\Delta \to \pi-\phi_\Delta$
inverts the signs of the energies is nicely
illustrated by comparing Fig.\ \ref{fig:flux_n0}
and Fig.\ \ref{fig:flux_n4}. The same relation is 
seen between the upper and the lower panels of Fig.\ \ref{fig:flux_n123}.
The middle panel of Fig.\ \ref{fig:flux_n123} remains
unchanged under inversion of the signs, which implies
that the upper and lower bands differ only by their
sign and that the middle band is identical to zero.

\begin{figure}[htb]
\begin{center}
	\includegraphics[width=0.8\columnwidth,clip]{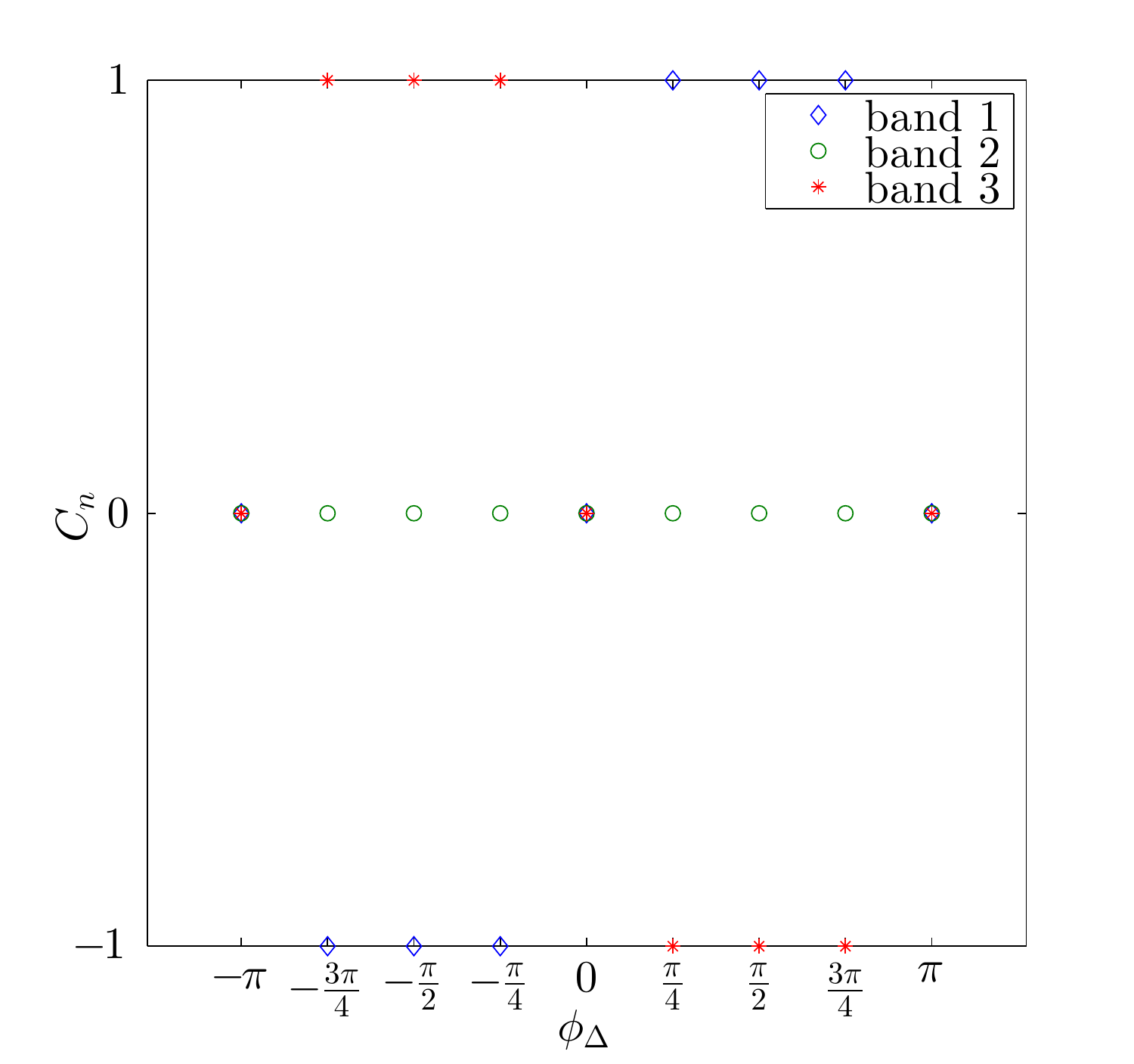}
\end{center}
\caption{(Color online) 
Chern number $C_n$ as defined in Eqs. \eqref{eq:chernzahl} computed
for various fluxes $\phi_\Delta$ for 
$\vartheta_1= \vartheta_2=\phi_\Delta/3$. 
The bands are numbered in order of ascending energy.
\label{fig:chern}}
\end{figure}

%Chern
Next, we turn to the Chern numbers of the bands. We did not
find a way to evaluate the expressions
in Eqs. \eqref{eq:chernzahl} analytically. Moreover, the two-dimensional
integrals are difficult to implement numerically to high precision.
But due to the robustness of the discrete Chern numbers the computation
of accumulated phases on a finite mesh is very accurate, even if
the mesh is not particularly dense \cite{fukui05}. Thus
we use this approach to compute the Chern numbers
shown in Fig.\ \ref{fig:chern}.

%%%%%%%%%%%%%%%%%%%%%%%%%%%%%%%%%%%%%
%%%%%%%%%%%%%%%%%%%%%%%%%%%%%%%%%%%%%%%%%%%%%%%%%%%%%%%%%%%%%%%%%%%%%%%%%%%%%%%

Several interesting observations can be made. First, the fluxes,  which
are multiples of $\pi$, indeed yield trivial bands. Second, incrementing
the flux by $\pi$ yields a swap of the Chern numbers due to
the inverted global sign of the energies. The change 
$\phi_\Delta \to \pi-\phi_\Delta$ leaves the Chern numbers unchanged.
Third, swapping the sign of the fluxes swaps also the sign of the 
Chern numbers. Fourth, the gradual evolution
of the bands from $\phi_\Delta=\pi/4$ to $\phi_\Delta=3\pi/4$ with energetically
well-separated bands is reflected in constant Chern numbers. 

Further values of $\phi_\Delta$ do not need to be analyzed because
of the $2\pi$-periodicity of the Chern numbers as a function of $\phi_\Delta$.
Thus the pattern in Fig.\ \ref{fig:chern} will occur repeatedly.

\subsection{Robustness against local potentials}

In order to illustrate that the Chern numbers are robust against small changes
of the Hamiltonian we turn to the case with local potentials. Clearly,
equal changes of all potentials $V_A, V_B$, and $V_C$ do not affect the
Chern number at all because they simply add an identity matrix to 
$\mathcal{H}_\vektor{k}$ which shifts the eigenenergy but has no
impact on the eigenvectors and hence no impact on the Chern numbers;
see Eqs.\ \eqref{eq:chernzahl}.

\begin{figure}[htb]
\begin{center}
	\includegraphics[width=0.8\columnwidth,clip]{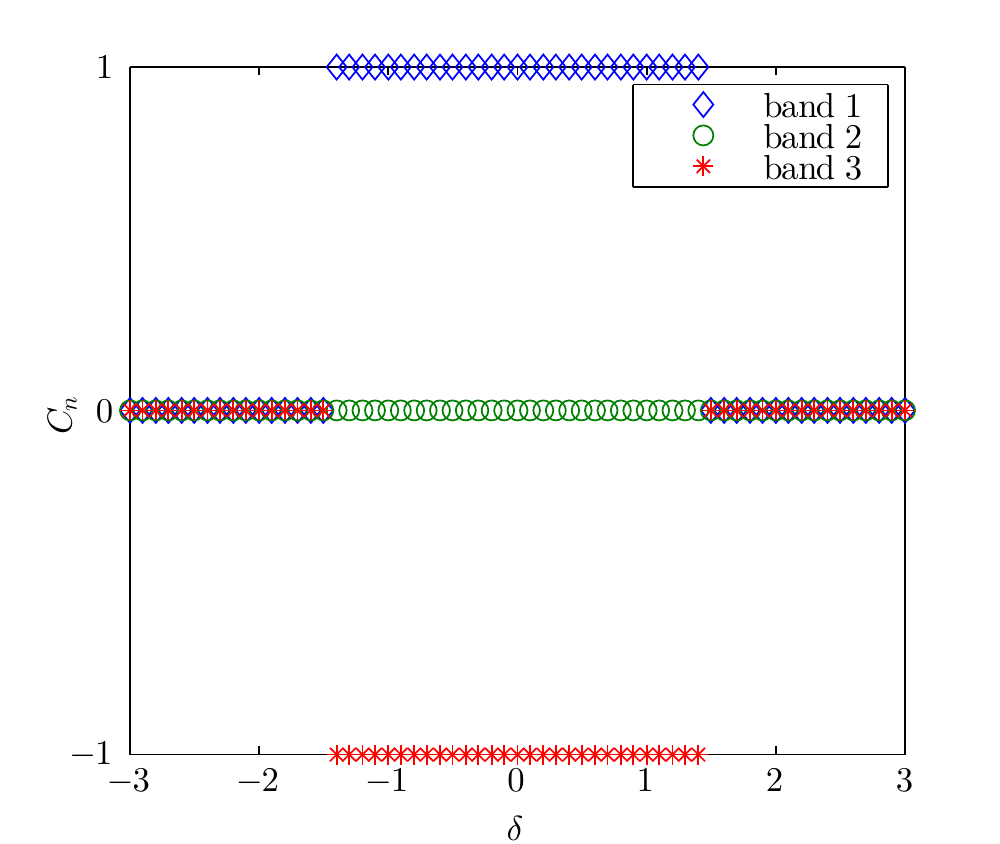}
\end{center}
\caption{(Color online) 
Chern number $C_n$ as defined in Eqs. \eqref{eq:chernzahl} computed
for flux $\phi_\Delta=\pi/2$ for $\vartheta_1= \vartheta_2=\pi/6$
at finite local potential according to $V_A=\delta=-V_B$
and $V_C=0$. The hopping $t$ is set to $1$ as before.
\label{fig:robust1}}
\end{figure}

\begin{figure}[htb]
\begin{center}
	\includegraphics[width=0.8\columnwidth,clip]{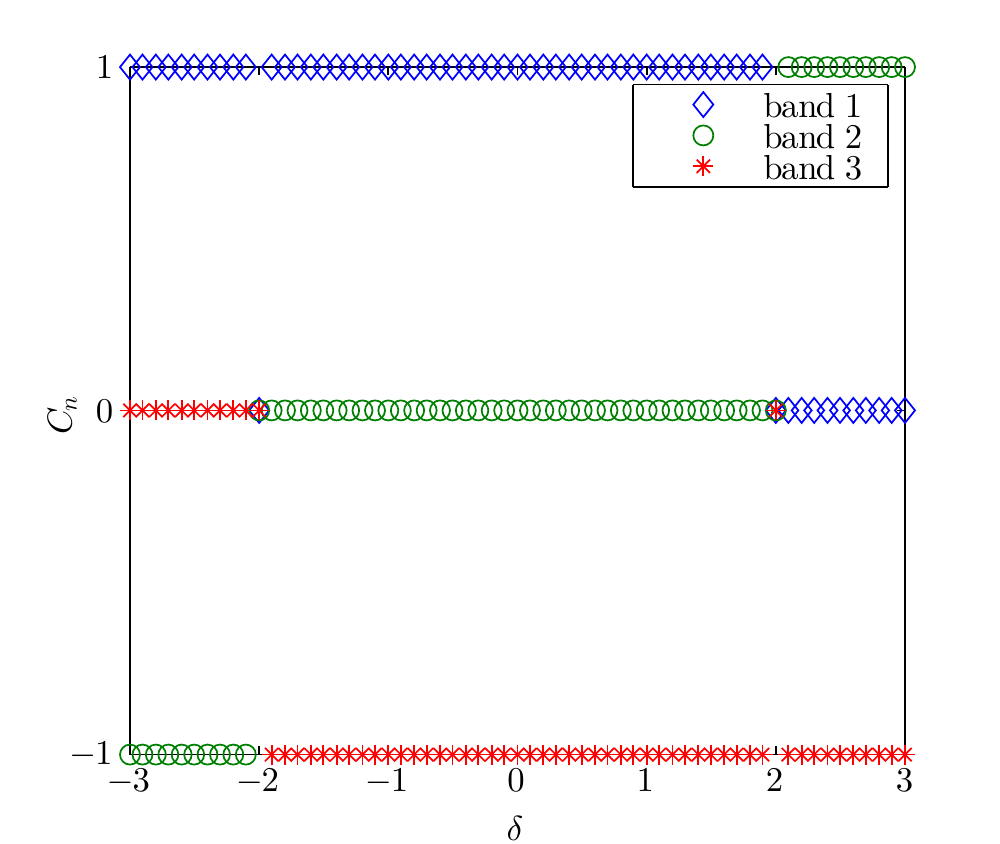}
\end{center}
\caption{(Color online) 
Chern number $C_n$ as defined in Eqs. \eqref{eq:chernzahl} computed
for flux $\phi_\Delta=\pi/2$ for $\vartheta_1= \vartheta_2 = \pi/6$
at finite local potential according to $V_A=\delta=V_C$
and $V_B=0$. The hopping $t$ is set to $1$ as before.
\label{fig:robust2}}
\end{figure}

For illustration, we study two nontrivial patterns of the
local potentials, namely $V_A=\delta=-V_B$ and $V_C=0$
in Fig.\ \ref{fig:robust1} and $V_A=\delta=V_C$
and $V_B=0$ in Fig.\ \ref{fig:robust2}. These patterns are applied
to the case $\phi_\Delta=\pi/2$ of which the dispersion without
local potentials is shown in the middle panel of Fig.\ \ref{fig:flux_n123}.
Clearly, the nontrivial Chern numbers of the lower and upper bands
are robust against the local potentials. It is necessary to apply
potentials of the order of $\delta\approx t$ to destroy the topological
phases because these phases are protected by gaps; see the middle panel of Fig.\ \ref{fig:flux_n123}. This is numerically shown in Figs.\ \ref{fig:robust1}
and \ref{fig:robust2} and can be analytically verified. For
the pattern $V_A=\delta=-V_B$ and $V_C=0$, the gaps close at $|\delta|=\sqrt{2}$,
and for the pattern $V_A=\delta=V_C$ and $V_B=0$, they close at $|\delta|={2}$.

These exemplary results illustrate the robustness of 
topological phases against perturbations. For the topological character
to be lost the bands must touch and become degenerate. 
A deformation alone is not sufficient.

%%%%%%%%%%%%%%%%%%%%%%%%%%%%%%%%%%%%%
%%%%%%%%%%%%%%%%%%%%%%%%%%%%%%%%%%%%%%%%%%%%%%%%%%%%%%%%%%%%%%%%%%%%%%%%%%%%%%%
%%%%%%%%%%%%%%%%%%%%%%%%%%%%%%%%%%%%%
%%%%%%%%%%%%%%%%%%%%%%%%%%%%%%%%%%%%%%%%%%%%%%%%%%%%%%%%%%%%%%%%%%%%%%%%%%%%%%%

\section{Edge states}

For measurable quantities the bulk properties are not of prime 
interest. The main difference between a usual, trivial band insulator
and a topological one occurs at the boundaries where states with
vanishing gap have to appear since otherwise the integer Chern numbers cannot
change from their finite value to zero.  These edge states are crucial
for the interesting properties of topologically ordered systems.

%%%%%%%%%%%%%%%%%%%%%%%%%%%%%%%%%%%%%
\begin{figure}[htb]
\begin{center}
	\includegraphics[width=0.9\columnwidth,clip]{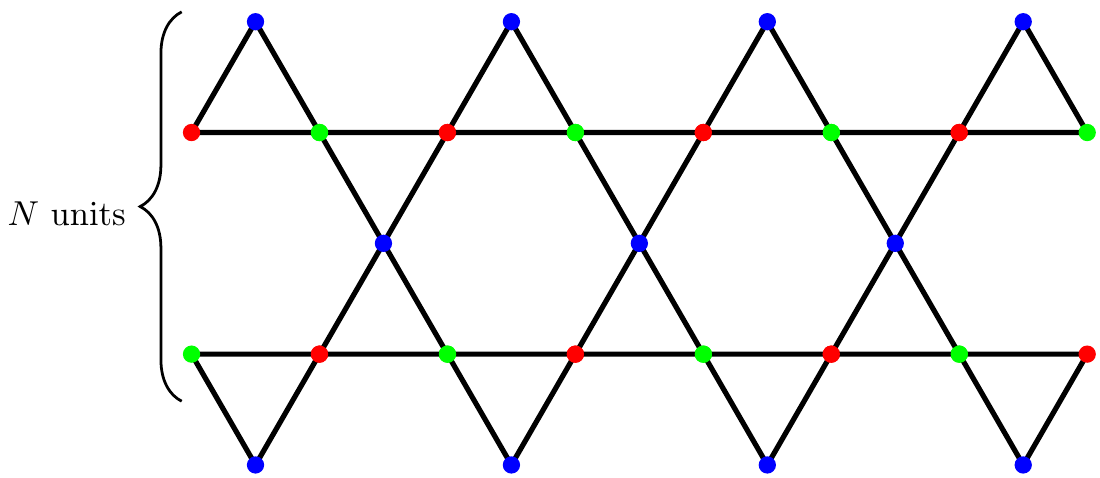}
	\includegraphics[width=0.9\columnwidth,clip]{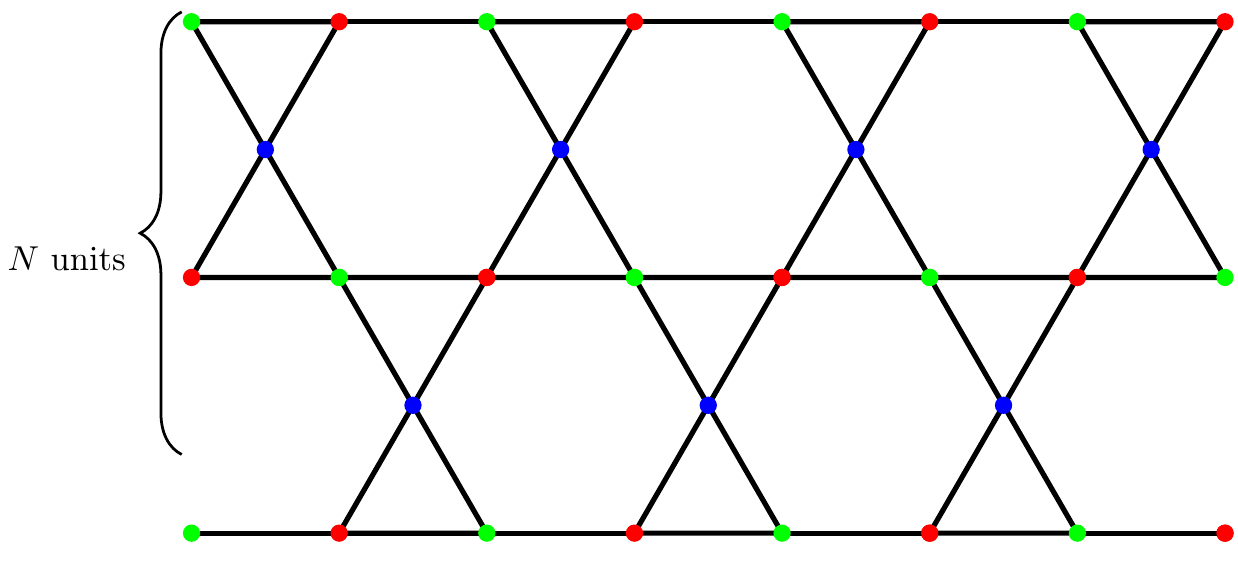}
	\includegraphics[width=0.9\columnwidth,clip]{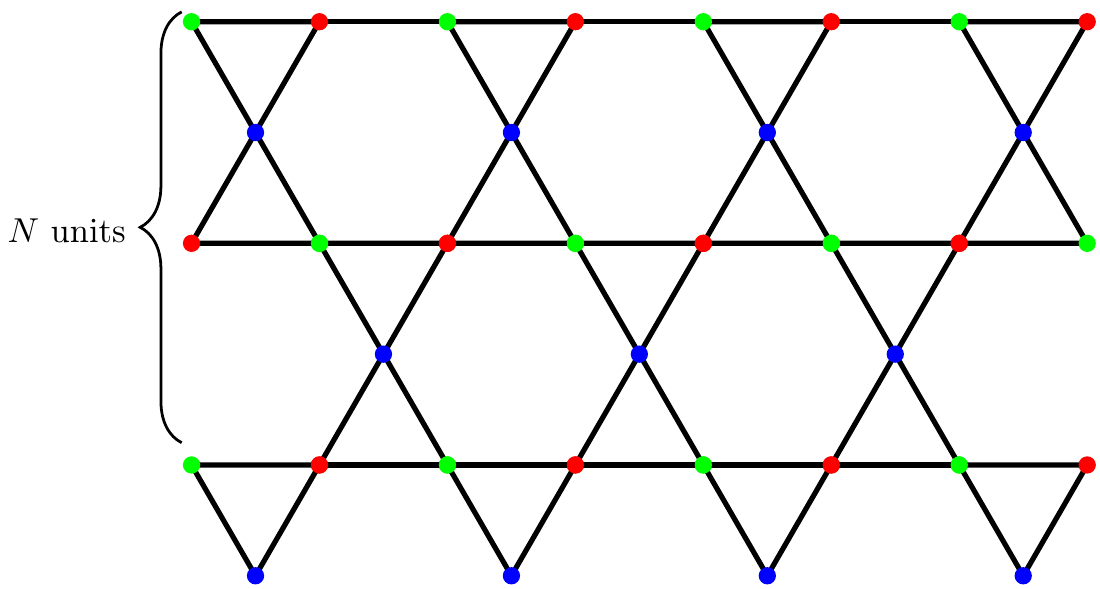}	
\end{center}
\caption{(Color online) 
Three types of edges: with triangular teeth at both edges (top),
for smooth boundaries without triangular teeth at both edges (middle),
and for mixed boundaries which are smooth at the upper edge
and saw-toothed at the lower edge (bottom). The bracket labeled ``$N$ units''
stands for $N$ times repeated units between the boundaries.
\label{fig:raender}}
\end{figure}
%%%%%%%%%%%%%%%%%%%%%%%%%%%%%%%%%%%%%%%%%%%%%%%%%%%%%%%%%%%%%%%%%%%%%%%%%%%%%%%

%%%%%%%%%%%%%%%%%%%%%%%%%%%%%%%%%%%%%
\begin{figure}[htb]
\begin{center}
	\includegraphics[width=0.8\columnwidth,clip]{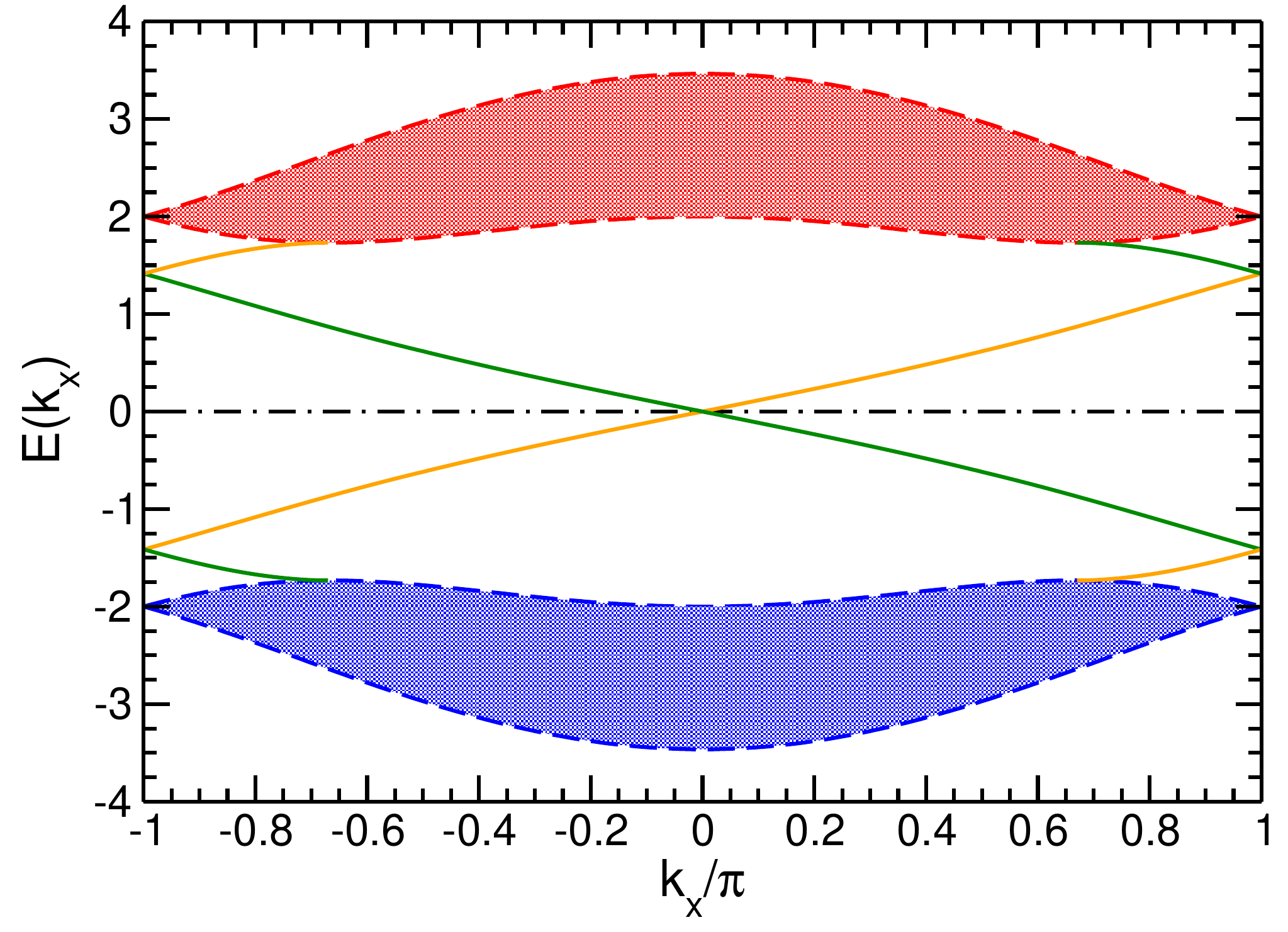}
	\includegraphics[width=0.8\columnwidth,clip]{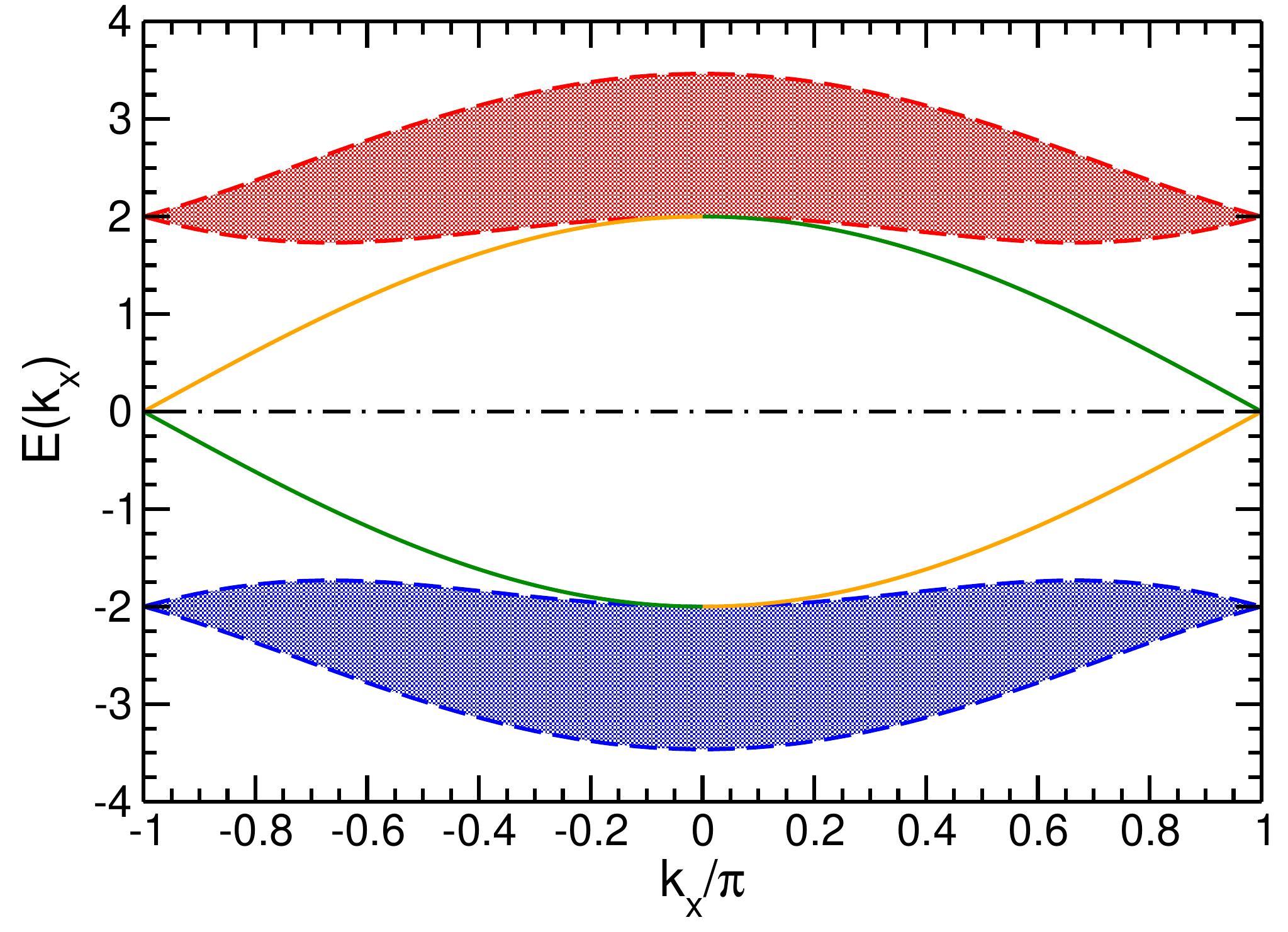}
	\includegraphics[width=0.8\columnwidth,clip]{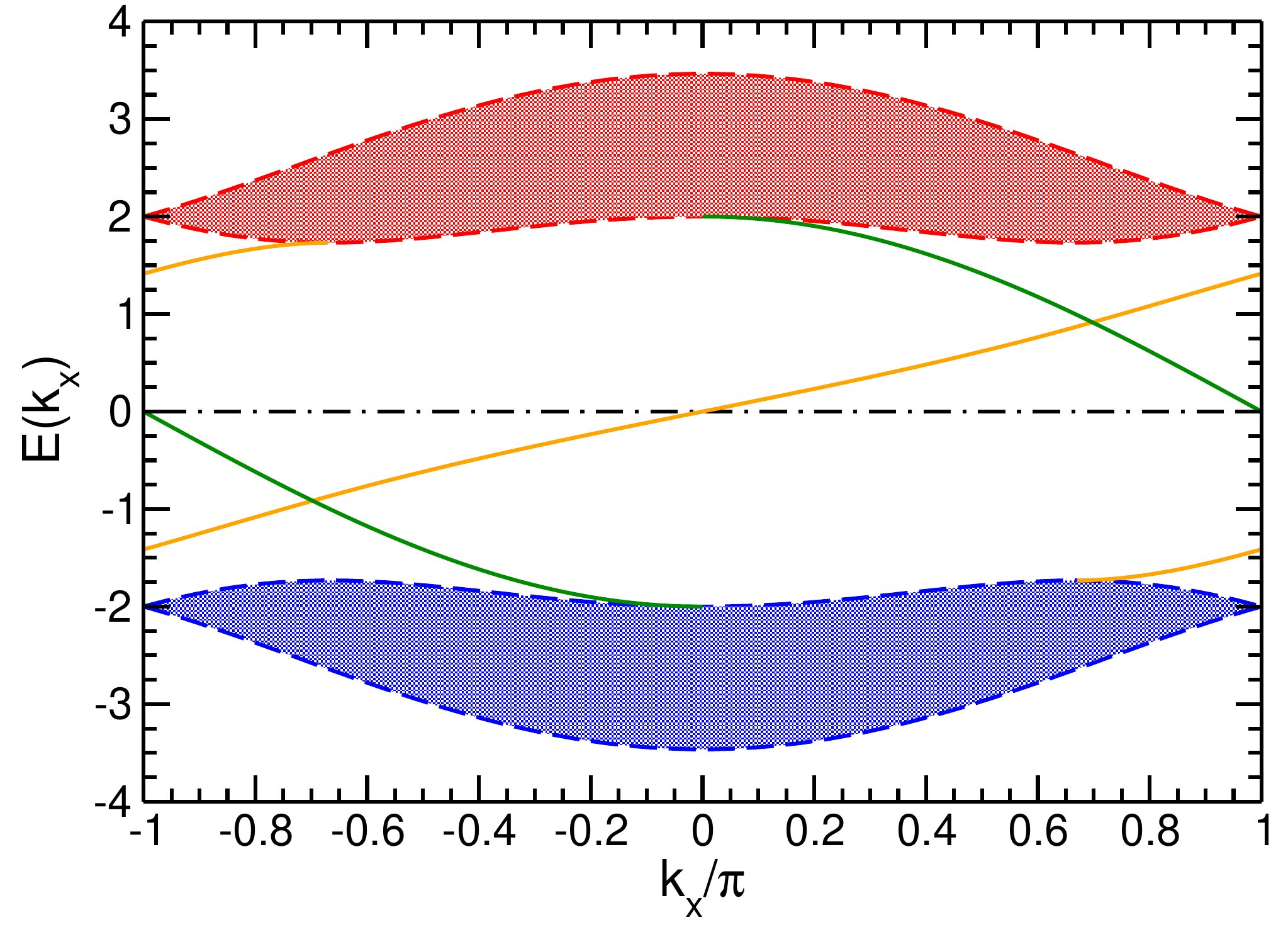}	
\end{center}
\caption{(Color online) 
Dispersions of the edge states at $\phi_\Delta=\pi/2$ 
if the edges are shaped as shown
in Fig.\ \ref{fig:raender}: the top, middle, and bottom panels here shows
the data relevant for the corresponding panel in Fig. \ref{fig:raender}.
The right-moving states from the lower edge 
are shown in orange and the left-moving ones from the upper edge 
in dark green. The shaded area stands for the continua
of the bulk states if $k_y$ is not fixed. The dashed line at zero
stands for the nondispersive bulk states of the trivial middle band;
cf.\ middle panel in Fig.\ \ref{fig:flux_n123}.
\label{fig:edge-dispersion}}
\end{figure}
%%%%%%%%%%%%%%%%%%%%%%%%%%%%%%%%%%%%%%%%%%%%%%%%%%%%%%%%%%%%%%%%%%%%%%%%%%%%%%%

To illustrate the existence of such edge states also in the kagome
lattice in a magnetic field we diagonalize strips of it. A strip is
infinitely extended in the $x$ direction such that $k_x$ is a conserved
quantum numbers. But in the $y$ direction the strip is of finite
extension. Three different situations of the boundaries are
displayed in Fig.\ \ref{fig:raender}. The height of the strips
is fixed by the number $N$ of repeated units of six sites
in the perpendicular $y$ direction.
The dispersions of the in-gap edges states converge very quickly 
for $N\to\infty$. Our results are based on computations with $N=80$
which amounts to the diagonalization of matrices of dimension 480.

The resulting dispersions are shown in Fig.\ \ref{fig:edge-dispersion}.
Strikingly, there occur significant differences for the dispersions of
the differently shaped boundaries. Even the group velocities
differ by about a factor of 2.

These differences make the assignment of the edge states to 
the upper and lower edges
particularly simple by analyzing mixed boundaries; see the lower panels
in Figs.\ \ref{fig:raender} and \ref{fig:edge-dispersion}.
Clearly, we partly find the dispersion from the upper panel 
for the right-moving states and from the lower panel for the left-moving states.
Thus the right-moving states live at the lower edge and the 
left-moving states at the upper edge. Of course, this will be swapped
if the sign of the magnetic field is inverted.

We conclude that  the shapes of the boundaries open
an interesting field for tuning the properties, for instance
the transport properties, of topologically nontrivial bands.

\section{Conclusions}

We have shown that a crystal lattice with basis equally
allows us to induce topologically nontrivial bands upon
application of a uniform magnetic field. Interestingly, 
the number of bands does not need to be increased
from the values at zero magnetic field, which is in contrast
to the situation in Bravais lattices \cite{hofst76,aidel15}.

%%%%%%%%%%%%%%%%%%%%%%%%%%%%%%%%%%%%%
\begin{figure}[htb]
\begin{center}
	\includegraphics[width=1.0\columnwidth,clip]{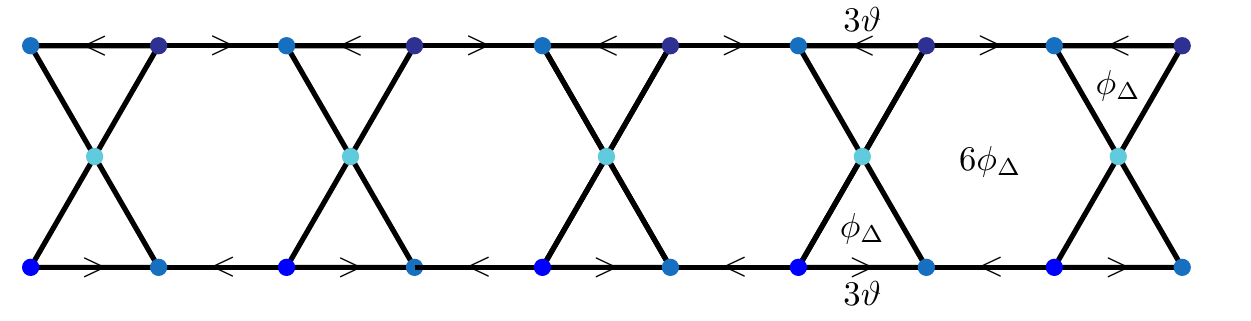}
\end{center}
\caption{(Color online) 
Phases after the gauge transformation $c_i^\dag\to c_i^\dag\exp(\mp \iu\vartheta)$
for A and B sites, respectively. The experimental realization should
be easier in this gauge.
\label{fig:gauge_exp}}
\end{figure}
%%%%%%%%%%%%%%%%%%%%%%%%%%%%%%%%%%%%%%%%%%%%%%%%%%%%%%%%%%%%%%%%%%%%%%%%%%%%%%%

In particular, we studied nearest-neighbor hopping 
on the kagome lattice with a three-fold
basis. This lattice is made of triangular loops, i.e., loops
with the smallest number of odd bonds which generically
induces frustration effects. Without magnetic field
the bands are not separated, but the special magnetic fields
which preserve the commutation of the translations induce
a proper splitting and nonzero Chern numbers for two
of the three bands. 
Interestingly, we found that the kagome
lattice in a uniform magnetic field of particular strengths
realizes Haldane models
on this lattice for particular quantized fluxes.

It would be interesting to realize crystal lattices
with nontrivial gauges experimentally \cite{jimen12,aidel15}.
In order not to be forced to induce finite phases 
$\vartheta:=\vartheta_1=\vartheta_2$ for all
hopping processes, we point out that a simple gauge transformation
eliminates all phases on bonds to or from sites of the sublattice
C. It consists of the additional factor $c_i^\dag\to c_i^\dag\exp(-i\vartheta)$
for A sites and  the additional factor $c_i^\dag\to c_i^\dag\exp(i\vartheta)$
for B sites. Of course, the fluxes through the triangles are left
unchanged because the phases on bonds between the A and B sites are tripled,
$\vartheta\to3\vartheta$, as shown in Fig.\ \ref{fig:gauge_exp}.

Finally, we investigated the edge states for three ways to cut the kagome
lattice along the same direction. Again, this possibility
is a particular feature for lattices with a basis. 
Surprisingly, we found a significant qualitative 
dependence of the dispersion of the edge states on the nature of the
boundaries. The shape and the position of the dispersions differ as 
well as the group velocities. Still there are left- and right-moving
modes which can be clearly assigned to the upper or lower edge.
The situation is particularly clear if the upper and lower
edges differ. Then we obtained two differing edge states, one
reflecting the behavior at the upper edge and one the behavior
at the lower edge. 

Experimental verification of this interesting
scenario is called for. One may envisage fascinating applications
because the transport properties will differ depending on the direction
of motion of a fermion from left to right or vice versa.
It is conceivable to build diode like devices if the properties
of the edges are tuned to maximize their differing characteristics,
for instance the group velocity. 

At present, the most promising systems for experimental realization
are not yet clear \cite{bergh13}. But recent advances in realizing
artificial gauges in systems of optical lattices \cite{jimen12,aidel15,jotzu14}
make it likely that soon the theoretical results presented here
can be put to experimental tests. The realization of precisely defined
edges may pose a problem, but any reproducible difference
between two edges of a strip of lattice will make a verification
possible.

Alternatively, various solid-state systems 
may constitute alternative routes towards experimental realizations,
for instance thin slabs of ferromagnetic Chern insulators
\cite{chang13,kou14,chang15}, honeycomb lattices built from atoms with higher
 atomic number \cite{wu14}, artificial lattices on surfaces \cite{krash11,han15}, 
or special lattices built from semiconductor nanostructures \cite{lan12}.
In any case, we conclude that designed 
crystal lattices constitute a promising field of research.

\begin{acknowledgments} 
We gratefully acknowledge helpful
input from Maik Malki and financial support of the Helmholtz Virtual Institute 
``New states of matter and their excitations.'' 
\end{acknowledgments}

\begin{appendix}

\section{Arbitrary magnetic flux}
\label{app:arbit}

\begin{figure}[htb]
\begin{center}
	\includegraphics[width=0.88\columnwidth,clip]{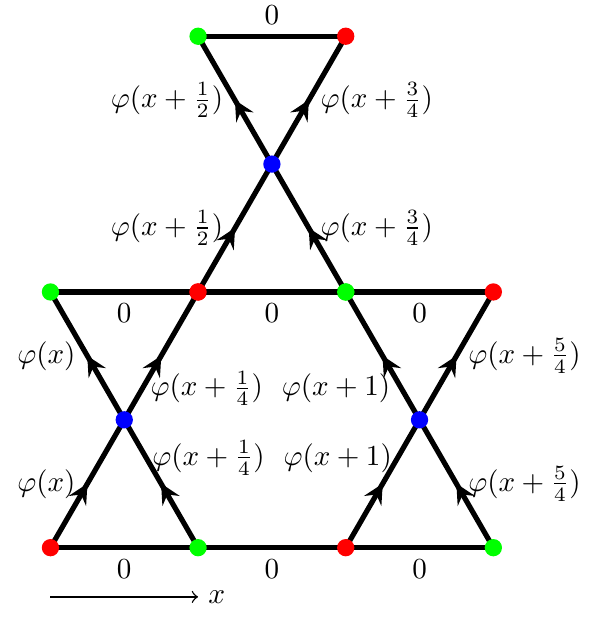}
\end{center}
\caption{(Color online) 
Gauge for arbitrary magnetic flux, i.e., for arbitrary homogeneous
magnetic field. Recall that the lattice constant, i.e., double
the distance between neighboring sites, is set to unity.
\label{fig:arbit}}
\end{figure}

In Fig.\ \ref{fig:arbit} a general gauge
is shown which is a lattice version of the Landau gauge.
The phases depend on the $x$ coordinate of their bond,  i.e.,
on the $x$ coordinate of the midpoint of their bond. The dependence
is linear,
\begin{equation}
\label{eq:gauge_real}
\varphi(x):= 4\phi_\Delta x,
\end{equation}
where we imply that the lattice constant is set to unity.

%%%%%%%%%%%%%%%%%%%%%%%%%%%%%%%%%%%%%
%%%%%%%%%%%%%%%%%%%%%%%%%%%%%%%%%%%%%%%%%%%%%%%%%%%%%%%%%%%%%%%%%%%

Then it is easy to see that indeed the flux through a triangle
is given by 
\begin{equation}
\phi_\Delta = \varphi(x+\frac{1}{4}) - \varphi(x)
\end{equation}
so that our notation is consistent. Similarly, we find
for the flux through the hexagon
\begin{subequations}
\begin{eqnarray}
\phi_{\mathrm{hexagon}} &=& 2\varphi(x+1) - 2\varphi(x+\frac{1}{4})\\
&=& 6\phi_\Delta
\end{eqnarray}
\end{subequations}
as it has to be for a homogeneous magnetic field.
In total, the flux through a unit cell is given by $8\phi_\Delta$.
Note that these results do not change if we consider shifted
triangles or hexagons in other parts of the lattice.
Shifts in the $y$ direction do not change the phases \eqref{eq:gauge_real} 
at all. Shifts in the $x$ direction do not change the fluxes because 
they depend on phase differences being independent of $x$
due to the linearity of Eq. \eqref{eq:gauge_real}.
So the gauge shown in Fig.\ \ref{fig:arbit} correctly describes 
the effect of an arbitrary homogeneous magnetic field perpendicular
to the plane of the lattice.

\section{Translational invariant gauge for specific magnetic fluxes}
\label{app:specific}

\begin{figure}[htb]
\begin{center}
	\includegraphics[width=0.78\columnwidth,clip]{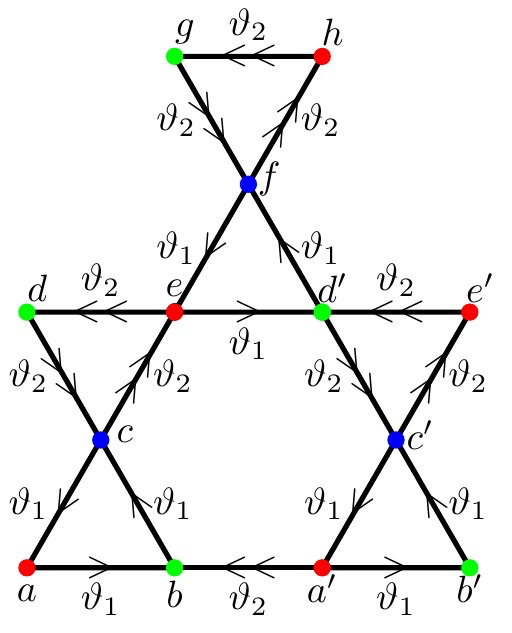}
\end{center}
\caption{(Color online) 
Gauge for specific magnetic fluxes, i.e., for specific homogeneous
magnetic fields. The necessary regauging on the various sites $a$
to $h$ and $a'$ to $e'$ are given in the main text.
\label{fig:specific}}
\end{figure}

Here we show that for $\phi_\Delta=n\pi/4$ the gauge in Fig.\ \ref{fig:arbit}
can be regauged to yield the gauge in Fig.\ \ref{fig:specific}.
For clarity we focus on the case $\vartheta_1=\vartheta_2=\phi_\Delta/3$,
i.e., $m_1=m_2=0$ in Eq.\ \eqref{eq:micros_relation}.

%%%%%%%%%%%%%%%%%%%%%%%%%%%%%%%%%%%%%
%%%%%%%%%%%%%%%%%%%%%%%%%%%%%%%%%%%%%%%%%%%%%%%%%%%%%%%%%%%%%%%%%%%

We leave the fermion at site $a$ in Fig.\ \ref{fig:specific}
unchanged, but we transform the others according to
\begin{subequations}
\label{eq:regauge1}
\begin{eqnarray}
c^\dag_a &\to& c^\dag_a,\\
c^\dag_b &\to& c^\dag_b \exp(i\vartheta_1),\\
c^\dag_c &\to& c^\dag_c \exp(2i\vartheta_1-i\varphi(x+1/4)),\\
c^\dag_e &\to& c^\dag_e \exp(3i\vartheta_1-i2\varphi(x+1/4)),\\
c^\dag_d &\to& c^\dag_d \exp(i\vartheta_1-i\varphi(x)-i\varphi(x+1/4)).
\end{eqnarray}
\end{subequations}
The annihilation operators are transformed by the complex
conjugate phase factors. Obviously, the phase factors are chosen
such that the phases on the bonds $ab$, $bc$, $ce$, and $cd$
match those in Fig.\ \ref{fig:specific}. We have to check
the resulting phase factor on $ca$ and compute
\begin{subequations}
\begin{eqnarray}
\vartheta_{ca} &=& -2\vartheta_1+\varphi(x+1/4)-\varphi(x)\\
&=& -(2/3)\phi_\Delta + \phi_\Delta\\
&=& \phi_\Delta/3\\
&=& \vartheta_1,
\end{eqnarray}
\end{subequations}
which is what we wanted to have. Analogously, we confirm
for the phase on bond $ed$
\begin{subequations}
\begin{eqnarray}
\vartheta_{ed} &=& \vartheta_1-\varphi(x)-3\vartheta_1+\varphi(x+1/4)\\
&=& -(2/3)\phi_\Delta + \phi_\Delta\\
&=& \phi_\Delta/3\\
&=& \vartheta_2.
\end{eqnarray}
\end{subequations}

If we now pass from the sites $a$ to $e$ to the sites
 $a'$ to $e'$ we can use almost the same transformations
\eqref{eq:regauge1} because the phases on the bonds
are changed by multiples of $\pi$ due to 
\begin{subequations}
\begin{eqnarray}
\phi(\tilde x+1) &=& \phi(\tilde x)+4\phi_\Delta\\
	&=& \phi(\tilde x)+n\pi  
\end{eqnarray}
\end{subequations}
for any $\tilde x$.
This is so because $\phi_\Delta$ takes the particular values $n\pi/4 $
representing the crucial step in our argument.

If $n$ is even we may use exactly the same transformations.
If $n$ is odd we regauge the center site $c'$ by the factor
$-1$ and we are back to the phases between the 
sites $a$ through $e$
and use the regauge transformation \eqref{eq:regauge1}.
In addition, it is obvious that the phases
on the bonds $ba'$ and $ed'$ correspond to the ones in
Fig.\ \ref{fig:specific}.

For completeness, we state the regauge transformations
on sites $f$, $g$, and $h$,
\begin{subequations}
\label{eq:regauge2}
\begin{eqnarray}
c^\dag_f &\to& c^\dag_f \exp(2i\vartheta_1-3i\varphi(x)-4i\phi_\Delta),\\
c^\dag_g &\to& c^\dag_g \exp(i\vartheta_1-4i\varphi(x)-6i\phi_\Delta),\\
c^\dag_h &\to& c^\dag_h \exp(3i\vartheta_1-4i\varphi(x)-7i\phi_\Delta),
\label{eq:h-site}
\end{eqnarray}
\end{subequations}
because they are not directly deduced from the ones
on $a$ to $e$. Straightforward calculations verify that
these transforms yield the phase pattern shown in Fig.\
\ref{fig:specific}. The sites resulting from a shift
by one lattice constant to the right are again transformed
either exactly the same way as the sites $e$ to $h$ and $d'$
if $n$ is even. For odd $n$, the same transformations are used except for an
additional sign change on the shifted center site $f'$. 

In the next upper row the phase pattern for the row of sites $a$ to $e$
is repeated. Thus one can reuse  the transformations \eqref{eq:regauge1}
except that all the phases must be taken \emph{relative}
to the phase of the $h$ site [Eq. \eqref{eq:h-site}].

In an analogous fashion, one can implement regauge
transformations yielding $\vartheta_1\ne\vartheta_2$, but
they are slightly more complex because they comprise a certain
phase shift upon $x\to x+1$.

In conclusion, we can regauge the case of a general homogeneous
magnetic field to the phases shown in Figs.\ \ref{fig:phases}
or \ref{fig:specific}.

\end{appendix}

%\bibliographystyle{apsrev}
%\bibliography{../../bibinput/liter10} 

\end{document}